\documentclass{iau} 
\usepackage{graphicx}
\usepackage{color}
\usepackage{url}
\usepackage{ulem}

\title[Spatially resolved study of the Local Group galaxies]
{Spatially resolved study on star and planet formation in Local Group
galaxies}

\author[Chikako Yasui] 
{Chikako Yasui$^1$}

\affiliation{$^1$National Astronomical Observatory of Japan, \\
2-21-1 Osawa, Mitaka, Tokyo, 181-8588, JAPAN \\
email: {\tt ck.yasui@gmail.com}}

\pubyear{2018}
\volume{347}  
\jname{Early Science with ELTs (EASE)}
\editors{N. Przybilla, K. Pollard \& A. Calamida, eds.}
\begin{document}

\maketitle

\begin{abstract}

We have been investigating the metallicity dependence of star and planet
 formation, particularly focusing on the outer Galaxy
with a Galactocentric distance ($R_g$) of $\gtrsim$15\,kpc, 
where the metallicity is determined to be as low as $\sim$$-$1\,dex.
We have obtained near-infrared (NIR) images of young clusters in the
outer Galaxy through the 8.2-m Subaru Telescope and have clearly
resolved the cluster members with mass detection limits of
$\sim$0.1\,$M_\odot$.
Consequently, we determined that the initial mass function (IMF) in the
outer Galaxy is consistent with that in the solar neighborhood 
with regard to high-mass slope and IMF peak.
Meanwhile, we suggested that  the lifetime of protoplanetary disks
is significantly shorter than that in the solar neighborhood.
We also reported a metallicity dependence of the disk lifetime. 
Future observations with higher spatial resolution and sensitivity by
using ELT will allow us to extend the spatially resolved studies on the
IMF and protoplanetary disk evolution to Local Group galaxies.
With high spatial resolution and sensitivity (i.e. 0.02\,arcsec and
$K\sim27$\,mag with an adaptive optics), stars with mass of down to
$\sim$0.1\,$M_\odot$ can be detected and also sufficiently resolved.
Based on such a study in a wider variety of environments, 
we can gain new insights related to environmental
effects of star and planet formation.

 
 \end{abstract}

\firstsection 
\section{Introduction}

We are investigating the environmental effects of star and planet
formation. Particularly, we have been focusing on metallicity
dependence.
Metallicity, or the abundance of metal, is known to increase with the
cosmic evolution owing to the element synthesis of stars and supernovae.
Even presently, only 2\,\% (in mass) of baryon exists in our solar system.
Nevertheless, metallicity is believe to be one of the most critical
factors for star and planet formation because dust is necessary to form
planet cores despite the very small mass fraction in disks ($\sim$1\,\%)
and also because metal is sensitive to heating and cooling in
star-forming processes and is directly related to radiative transfer.
If any dependences are observed, these can have a strong influence on 
theories of star and planet formation 
(e.g., \cite[Elmegreen et al. 2008]{Elmegreen2008}, \cite[Ercolano \&
Clarke 2010]{Ercolano2010}).

\section{Science Cases}

In the investigation of metallicity dependence of star and planet formation, we are
focusing on the IMF and the lifetime of protoplanetary disks.
Considering that stars are fundamental components in galaxies and that
stellar mass determines their evolutionary path, the IMF is one of the most
fundamental parameters that determines the physical and chemical evolution
of stellar systems.
Because planets are formed in protoplanetary disks and disk duration
can regulate planet formation, disk lifetime is one of the most
fundamental parameters directly connected to the planet formation
probability.
Whether the IMFs and disk lifetime derived in the solar neighborhood are
universal or sensitive to environmental conditions is under debate
(\cite[Lada \& Lada 2003]{Lada2003}).
In both these science cases, resolved study is crucial for avoiding
blending/stochastic effects.

Our observational targets are star-forming clusters that
satisfy the following two criteria: 
i) young star-forming regions ($<$5Myr old) because massive stars still exist, 
age spreads are small, and most of the protoplanetary disks are likely present and 
ii) moderate-scale clusters, with a cluster mass ($M_{\rm cl}$) of
$\sim$$10^2$--$10^3$\,$M_\odot$. 
In the solar neighborhood, $M_{\rm cl}$ of  $\sim$$10^2$\,$M_\odot$ 
class is most common (\cite[Adams et al. 2006]{Adams2006}).
Moreover, even the largest star-forming region in the solar neighborhood,
that is, Orion Nebula Cluster (ONC; 
Fig.~\ref{fig:ONC}, left), has $M_{\rm cl}$ of $\sim$1000\,$M_\odot$
(\cite[Lada \& Lada 2003]{Lada2003}).
In the derivation of the absolute metallicity dependence, we focus on
the young clusters with $M_{\rm cl}$ of $<$$10^4$\,$M_\odot$ to combine
the results with that of nearby clusters.
Indeed, non-standard IMF is indicated for starburst clusters ($M_{\rm
cl}$ of $>$$10^4$\,$M_\odot$; e.g., Arches, Westerlund 1, NGC 3603;
\cite[Bastian, Covey, \& Meyer 2010]{Bastian2010}) although it may be
owing to blending effects.
Other possible environmental effects, aside from than 
metallicity, are suggested on disk lifetime (e.g., cluster mass and
cluster density; references in \cite[Yasui et al. 2016a]{Yasui2016a}).
Note that the solar system is also most likely to be formed in 
 $M_{\rm cl} \sim 10^3$\,$M_\odot$ class cluster (e.g., \cite[Adams
 2010]{Adams2010}).

 \begin{table}
  \begin{center}
   \caption{Distances of target clusters vs. their stellar separation.}
  \label{tab1}
 {\scriptsize
  \begin{tabular}{|l|c|c|c|c|}\hline 
   {\bf Distances} & {\bf Examples}
  & {\bf Stellar separation} & {\bf Comments} \\
   \hline
   10 (pc) & & \\ \hline 
   2 (kpc) & Solar neighborhood & 10$''$ \\ \hline 
   10 (kpc) & Our Galaxy & 2$''$\\ \hline 
   50 (kpc) & LMC and SMC & 0.4$''$ & Current limit \\ \hline 
   100 (kpc) &  & 0.2$''$\\ \hline 
   500 (kpc) &  & 0.04$''$\\ \hline 
   1 (Mpc) & Local Group & 0.02$''$ & Limit of ELTs \\ \hline  
   5 (Mpc) &  & 0.004$''$ & Limit of 100\,m telescopes \\ \hline
   10 (Mpc) & & 0.002$''$\\ \hline  
  \end{tabular}
  }
 \end{center}
\vspace{1mm}
 \scriptsize{
 {\it Notes:} 
Assumed typical stellar separation for nearby clusters $\sim$0.1\,pc
  (\cite[Adams et al. 2006]{Adams2006}). 
 }
 \end{table}

\section{Pre-studies Prior to ELTs}

{\underline{\it 8--10\,m class telescopes}}. 
The distances to the target clusters and their stellar separations are
summarized in Table~\ref{tab1}.
In this study, we assumed that the typical stellar separation of nearby
cluster members is $\sim$0.1\,pc, which is estimated from young embedded
clusters in the solar neighborhood (\cite[Adams et
al. 2006]{Adams2006}).
The distance of $\sim$50--100\,kpc (LMC and SMC) is the current limit
for spatially resolved studies that used 8--10\,m class telescopes with
an adaptive optics (AO) that assures decent Strehl ratio ($\sim$0.1)
even for extragalactic objects.

As  a first step,
we are focused on the outer Galaxy, which we defined as a region with
$R_g$ of $\gtrsim$15\,kpc
because it is known to have a metallicity of as low as  $\sim$$-$1\,dex
despite its proximity, in comparison with the galaxies in the local universe. 
The environments are similar to that in nearby dwarf galaxies, damped 
Lyman-alpha systems, and that in the early stage of the the
Galactic disk formation (e.g., \cite[Ferguson et al. 1998]{Ferguson1998},
\cite[Kobayashi et al. 2008]{Kobayashi2008}).
We have obtained NIR images of more than 10 star-forming
clusters in the outer Galaxy through the 8.2-m Subaru Telescope (e.g.,
Yasui et al. 2008, 2010, 2016b, 2016c).
%
High spatial resolution (i.e. $\lesssim$0.5$''$) and relatively small
distances to the clusters enabled us to resolve individual cluster
members in each cluster even without AO.
Hence, the limiting magnitudes of $K_S \sim 18$--21\,mag (10$\sigma$)
are achieved. This magnitude corresponds to the mass of
$\sim$0.1\,$M_\odot$, which is much smaller than that achieved in nearby
dwarf galaxies, such as LMC and SMC ($\sim$1\,$M_\odot$).
Based on the fitting of K-band luminosity functions, we determined that
the IMF in the outer Galaxy is consistent with that in the solar
neighborhood with regard to the high-mass slope and IMF peak
($\sim$0.3\,$M_\odot$; Yasui et al. 2017, 2008),
suggesting that the IMF down to substellar mass regime
($\sim$0.1\,$M_\odot$) have no dependence on the metallicity down to
$\sim$$-$1\,dex.
However, we identified that the fraction of stars with a K-band excess
(originated from the inner circumstellar dust disk at a radii of $r \le
0.1$\,AU) is significantly lower than that in the solar neighborhood
(\cite[Yasui et al. 2009]{Yasui2009}), suggesting a metallicity
dependence of the disk lifetime (\cite[Yasui et al. 2010]{Yasui2010}).

{\underline{\it JWST}}.
We are planning on the NIR and mid-infrared (MIR) imaging with
 MIRI/NIRCam as a part of the Guaranteed Time Observation (GTO 1237; PI:
 Michael Ressler (JPL)).
The expected limiting magnitudes with 30\,min integration (Vega,
5$\sigma$) are $\sim$23\,mag at 4.4\,$\mu$m and 17.3\,mag at
12.8\,$\mu$m, whereas the spatial resolution will be $\sim$0.4$''$ at
10\,$\mu$m.
The sensitivities are significantly higher compared with existing
telescopes (Fig.~\ref{fig:sensitivity}; Thirty Meter Telescope (TMT)
Planning instructions (2012):
\url{https://tmt.nao.ac.jp/document/brochure/tmtbb-ols.pdf}) both in NIR
and MIR. However, the spatial resolution is comparable. 
Therefore, we will focus on the outer Galaxy for the JWST study.
The mass detection limit will be $\sim$8\,$M_J$ in NIR (Lyons/DUSTY
evolutionary models of \cite[Chabrier et al. 2000]{Chabrier2000};
cf. $\sim$0.1\,$M_\odot$ for LMC and SMC).
Thus, we can determine whether stars down to 0.5\,$M_\odot$ still have
disks in MIR (calculated based on the case of IC 348 at $D=320$\,pc by
\cite[Lada et al. 2006]{Lada2006}).
By using the JWST, IMF in this low-metallicity environment down to
substellar mass regime ($<$0.1\,$M_\odot$) can be derived.  The IMF in
the very low mass end can be a sensitive function of the formation
environment (\cite[Lada \& Lada 2003]{Lada2003}).
This study will be the first to research on brown dwarfs and planetary
mass objects in low-metallicity environments, enabling us to investigate
whether they are common in such environments.
We will also gain new insights related to the {\it outer}
($\sim$0.1--5\,AU) circumstellar disk evolution thanks to the high
sensitivities of JWST in MIR.
Hence, a comprehensive comparison of disk lifetimes in low-metallicity
regions with that in the solar neighborhood that has been extensively
characterized by {\it Spitzer} (e.g., \cite[Hern{\'a}ndez et
al. 2007]{Hernandez2007}) will be possible.

\section{Future Prospects with ELTs}

\begin{figure}
  \begin{minipage}{\hsize}
    \begin{minipage}{0.6\hsize}
      \hspace{-2.5em}
     \includegraphics[bb=0 0 2064 984, width=8.9cm]{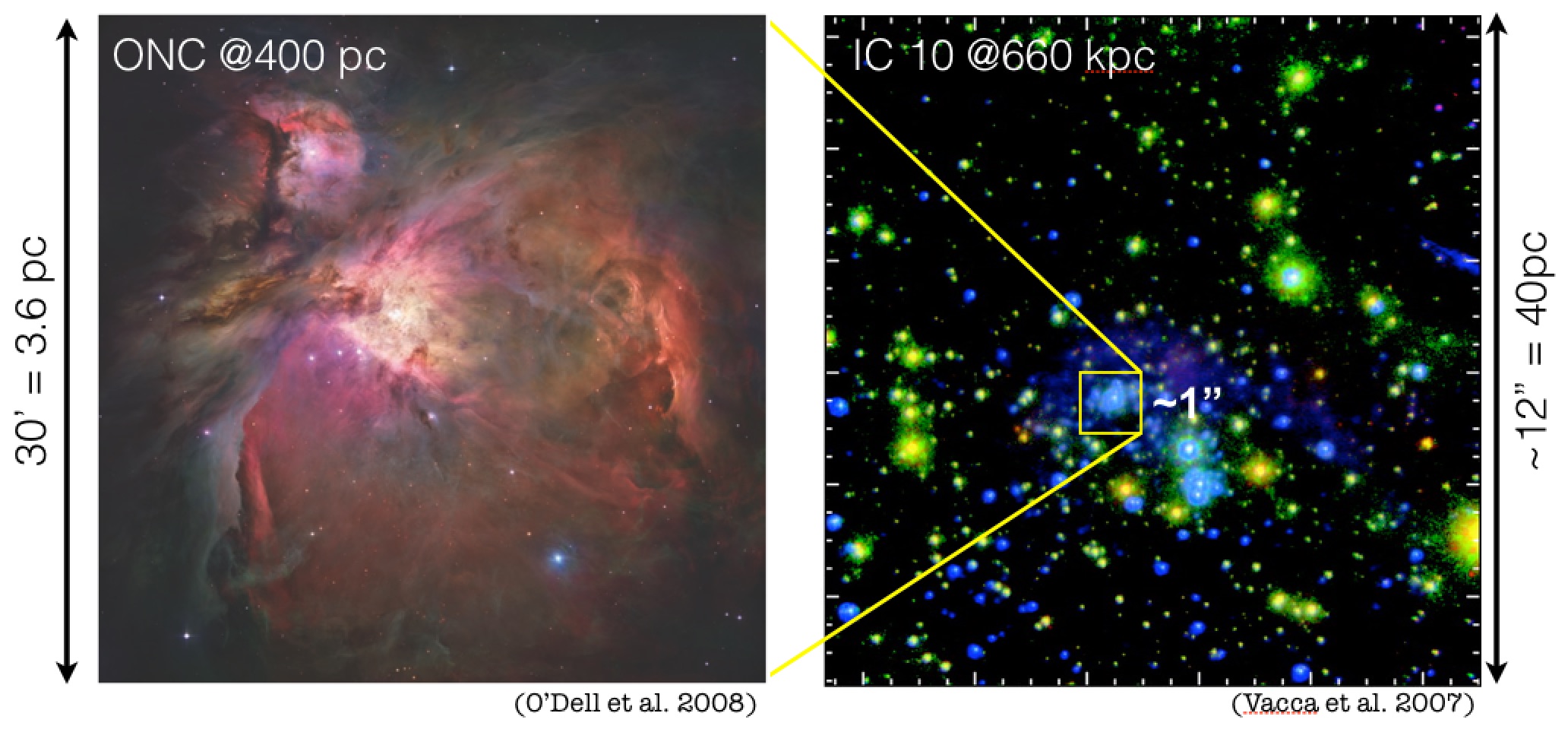}
     \caption{Comparison of the spatial coverage.  Even the ONC spans
     only $\sim$1$"$ at the extragalactic distances.} 
      \label{fig:ONC}
    \end{minipage}
    \begin{minipage}{0.4\hsize} 
	\includegraphics[bb=0 0 774 800, height=5cm]{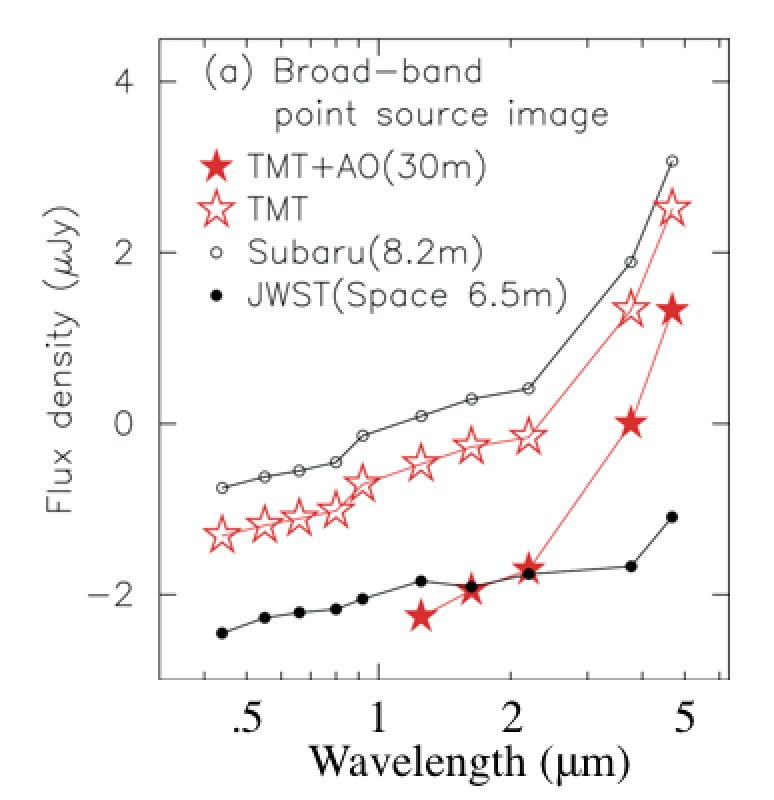}
	\caption{Comparison of the sensitivities.}
	\label{fig:sensitivity}
    \end{minipage}

  \end{minipage}
\end{figure}

ELTs offer significantly high sensitivities in $\lambda \lesssim
2$\,$\mu$m compared with existing telescopes, but they are comparable to
JWST in long NIR and lower MIR sensitivities
(Fig.~\ref{fig:sensitivity}).
In the case of NIR imaging with TMT/IRIS, limiting magnitudes with 5-h
integration (Vega, 10$\sigma$) will be 27.6\,mag at the $K$-band
(2.2\,$\mu$m) (\cite[Moore et al. 2014]{Moore2014}).
Meanwhile, the spatial resolution will be as small as $\sim$0.02$''$
with AO at the K-band, which is significantly higher than before,
enabling us to extend the spatially resolved studies to the Local Group
for the first time (Table~\ref{tab1}).
For 1\,Myr-old targets at $D=770$\,kpc, the mass detection limit of
$\sim$0.1\,$M_\odot$ (assuming that $A_V = 0$\,mag) /
$\sim$0.5\,$M_\odot$ (assuming that $A_V = 10$\,mag),
based on the isochrone model of \cite[Siess, Dufour, \& Forestini
(2000)]{Siess2000},
and stellar separation of 0.08\,pc, which is less than the typical stellar
separations in clusters (0.1pc; \cite[Adams et al. 2006]{Adams2006}), 
will be achieved.

The target star-forming regions are as follows: 
\begin{itemize}
	
 \item Spiral Galaxies (M31, M33): Many star-forming clusters exist.  We
       can derive the absolute metallicity dependence by observing the
       same type of galaxies as our Galaxy.

 \item Dwarf galaxies, e.g., IC 10 (Fig.~\ref{fig:ONC}, right), NGC
       6822, and Leo I. 
      We can explore wider metallicity range down to $\sim$$-$2\,dex.

\end{itemize}

The spatially resolved study on the Local Group galaxies should be one
of the most cutting-edge science themes with ELTs.
However, even ONC located at $D=400$\,pc (covering $\sim$$30'\times30'$,
corresponding to $\sim$$3.6\times3.6$\,pc$^2$), covers only $\sim$1$''$
square (Fig.~\ref{fig:ONC}) if it is located at $D=660$\,kpc (i.e. the
distance of IC 10).
Therefore, the achievement of the spatially resolved study on the Local
galaxy is heavily dependent on how effective the AO systems on the ELTs 
will work.


\begin{thebibliography}{}
 \bibitem[Adams et al.(2006)]{Adams2006}
 {Adams, F.C., Proszkow, E.M., Fatuzzo, M., and Myers, P.C.} 2006, \textit{ApJ}
 641, 504 

\bibitem[Adams(2010)]{Adams2010}
{Adams, F.~C.} 2010, \textit{ARAA}, 48, 47 

 \bibitem[Bastian, Covey, and Meyer(2010)]{Bastian2010}
{Bastian, N., Covey, K.R., \& Meyer, M.R.} 2010, \textit{ARAA}, 48, 339 

\bibitem[Chabrier et al.(2000)]{Chabrier2000}
 {Chabrier, G., Baraffe, I., Allard, F., \& Hauschildt, P.} 2000, \textit{ApJ},
 542, 464
 
 \bibitem[Elmegreen et al.(2008)]{Elmegreen2008} 
{Elmegreen, B.~G., Klessen, R.~S., Wilson, C.~D.} 2008, \textit{ApJ}, 681, 365

 \bibitem[Ercolano \& Clarke(2010)]{Ercolano2010}
{Ercolano, B., \& Clarke, C.J.} 2010, \textit{MNRAS}, 402 2735.

 \bibitem[Ferguson et al.(1998)]{Ferguson1998}
 {Ferguson, A.~M.~N., Gallagher, J.~S., Wyse, R.~F.~G.} 1998,
\textit{AJ}, 116, 673 

\bibitem[Hern{\'a}ndez et al.(2007)]{Hernandez2007}
{Hern{\'a}ndez J., et al.} 2007, \textit{ApJ}, 662, 1067 

\bibitem[Kobayashi et al.(2008)]{Kobayashi2008}
{Kobayashi, N., Yasui, C., Tokunaga, A.~T., \& Saito, M.} 2008,
\textit{ApJ}, 683, 178 

	 
 \bibitem[Lada \& Lada(2003)]{Lada2003}
{Lada, C.~J., \& Lada, E.~A.} 2003, \textit{ARAA}, 41, 57

\bibitem[Moore et al.(2014)]{Moore2014}
{Moore A.~M., et al.} 2014, SPIE, 9147, 914724 


		


 \bibitem[O'Dell et al.(2008)]{O'Dell2008}
O'Dell, C.~R., Muench, A., Smith, N., \& Zapata, L.\ 2008, Handbook of
Star Forming Regions, Volume I, 4, 544

		
\bibitem[Siess, Dufour, \& Forestini(2000)]{Siess2000}
{Siess, L., Dufour, E., \& Forestini, M.} 2000,
 {\it A\&A}, 358, 593

		
\bibitem[Vacca et al.(2007)]{Vacca2007}
Vacca, W.~D., Sheehy, C.~D., \& Graham, J.~R.\ 2007, {\it ApJ}, 662, 272 


		 
 \bibitem[Yasui et al.(2010)]{Yasui2010}
 {Yasui, C., Kobayashi, N., Tokunaga, A.T., Saito, M., \& Tokoku, C.}
 2010, \textit{ApJ} (Letters), \textit{723}, L113


 \bibitem[Yasui et al.(2009)]{Yasui2009}
 {Yasui, C., Kobayashi, N., Tokunaga, A.T., Saito, M., \& Tokoku, C.}
 2009, \textit{ApJ}, 705, 54

 \bibitem[Yasui et al.(2017)]{Yasui2017}
{Yasui C., Izumi N., Saito M., Kobayashi N.} 2017, IAUS, 321, 34 


 \bibitem[Yasui et al.(2008)]{Yasui2008}
 {Yasui C., Kobayashi N., Tokunaga A.~T., Terada H.,
 Saito M.} 2008, \textit{ApJ}, 675, 443
		 



 \bibitem[Yasui et al.(2016a)]{Yasui2016a}
{Yasui, C., Kobayashi, N., Hamano, S., Kondo, S., Izumi, N., Saito, M.,
\& Tokunaga, A.T.} 2016a, \textit{ApJ}, 817, 181

 \bibitem[Yasui et al.(2016b)]{Yasui2016b}
{Yasui C., Kobayashi N., Saito M., \& Izumi N.} 2016b, \textit{AJ}, 151, 115 

 \bibitem[Yasui et al.(2016c)]{Yasui2016c}
{Yasui C., Kobayashi N., Tokunaga A.~T., Saito M., \& Izumi N.} 2016c, 
\textit{AJ}, 151, 50 



\end{thebibliography}
\end{document}